\documentclass{elsart}

\usepackage{graphicx}

\begin{document}

\begin{frontmatter}

\title{Nonlocality and the Correlation of Measurement Bases}
\author{Daegene Song}
\address{Korea Institute for Advanced Study, Seoul 130-722, Korea}


\begin{abstract}
Nonlocal nature apparently shown in entanglement is one of the
most striking features of quantum theory. We examine the locality
assumption in Bell-type proofs for entangled qubits, i.e. the
outcome of a qubit at one end is independent of the basis choice
at the other end.  It has recently been claimed that in order to
properly incorporate the phenomenon of self-observation, the
Heisenberg picture with time going backwards provides a consistent
description. We show that, if this claim holds true, the
assumption in nonlocality proofs that basis choices at two ends
are independent of each other may no longer be true, and may pose
a threat to the validity of Bell-type proofs.

\end{abstract}


\end{frontmatter}

The nonlocal nature exhibited in quantum entanglement is arguably
the most distinctive departure from classical physics. After
Bell's pioneering work \cite{bell} in testing nonlocality in
entangled quantum systems, a number of variations of theoretical
models \cite{CHSH,GHZ,hardy} and experimental confirmations have
followed \cite{aspect}. While these truly marvelous results appear
to have confirmed the validity of quantum theory and triumphed
over locality imposed by relativity, subtle related issues remain.
That is, although quantum theory indeed appears to possess
nonlocality, this property cannot be used for superluminal
signalling. This is rather puzzling because there seems to exist
faster-than-light influencing yet superluminal signalling is not
allowed.  Another puzzling feature related to entanglement is the
negativity shown in the conditional entropy of entangled quantum
systems \cite{adami}. While a number of interpretations have been
made with regard to this negativity \cite{winter}, this issue is
still considered to be unsettled \cite{hayden2}.  Due to this
negativity, Cerf and Adami \cite{adami} have proposed to interpret
entanglement as a qubit and anti-qubit correlation where
anti-qubit is a qubit traveling backwards in time.

In Bell-type inequality proofs such as that of
Clauser-Horne-Shimony-Holt \cite{CHSH}, one of the critical
assumptions is the locality condition: that is, the outcome of a
particle, say at Alice's end, is independent of the basis choice
at the other end, or at Bob's end. This then implies that the
basis choice at one end cannot be correlated with the basis choice
at the other end. This is because if the basis choice were
correlated, i.e., if the basis choice at one end is influenced by
the basis choice at the other end, then a qubit at Bob's end could
learn about Alice's basis choice through the measurement basis at
Bob's end. The goal of this paper is to examine this particular
assumption made in Bell-type proofs, i.e., we wish to consider if
the assumption that basis choices at two ends are uncorrelated is
sound and solid. In order to examine this assumption in fuller
detail, let us review the concept of the physical reality defined
by Einstein, Podolsky, and Rosen (EPR) \cite{EPR} as follows:
\\
\\
{\it{If, without in any way disturbing a system, we can predict with
certainty (i.e., with probability equal to unity) the value of a
physical quantity, then there exists an element of physical reality
corresponding to this physical quantity}}.
\\
\\
Let us consider EPR's definition of physical reality with a single
qubit.  Let us denote the value of a physical quantity for a given
qubit as $N_Q$.  A measurement on the single qubit can be done with
a certain choice of observable and the outcome would be $\pm 1$. If
we denote the choice of observable as $N_C$, then the value of a
physical quantity, i.e., the physical reality for the qubit, can be
written as follows,
\begin{equation}
N_Q (N_C) = \pm 1
\end{equation}
That is, the physical reality of the qubit is obtained as a function
of the choice of measurement basis. We now wish to examine the
physical reality of the choice of measurement basis, i.e., $N_C$.
Certainly, there are a number of ambiguous elements if we wish to
discuss the physical reality of the basis choices.  For instance, it
is unclear what makes the decision to choose certain basis, i.e., is
it the computer random number generator, a detector, or a human
brain. Instead of considering the physical reality of basis choice
as a function of these ambiguous elements, we wish to consider the
$N_C$ as a function of the value of a physical quantity, i.e.,
$N_Q$. In doing so, there is a big assumption since the decision of
making an observable choice takes place prior to the eigenvalue
outcome, we are considering the physical reality of the past event
as a function of the future event.

In \cite{song1,song2}, the observables were considered as
observer's reference frame in observing and measuring a given
quantum state. In particular, it was argued that when it is the
unitary transformation of the observer's reference frame that is
observed by the same observer, both the Schr\"odinger and the
Heisenberg pictures cannot describe this phenomenon consistently.
It was then argued \cite{song1} that in order to correctly
describe the observation of observer's own reference frame is to
consider the Heisenberg picture with time going backwards, i.e.,
it is the observables, or the observer's reference frame, that is
evolving backwards in time. With this argument in \cite{song1},
let us consider a simple example. Suppose for a given qubit
$|0\rangle$, an observer is to measure this qubit either in $Z$-
or $X$-basis. If the observer chooses $Z$-basis, then he applies
an identity operator and measure with $\sigma_z = |0\rangle\langle
0| -|1\rangle\langle 1|$. When $X$-basis is chosen, the observer
applies a Hadamard gate, $H\equiv \frac{1}{2}(|0\rangle\langle 0|
+ |0\rangle\langle 1| + |1\rangle\langle 0| - |1\rangle\langle
1|)$, to the given qubit and measures $\sigma_z$ and the
eigenvalue outcome $\pm 1$ would be obtained.  In the Heisenberg
picture, rather than the state being unitarily transformed, it is
the basis choice that is evolving.  For convenience, suppose at
$t=0$, the observer chooses to measure either in $Z$- or
$X$-basis. Then either ${\bf{1}}$ or $H$ is applied to $Z$ basis
and $Z$ or $HZH =X$ is obtained. Since we are taking an assumption
that the unitary transformation is being done in time backward
manner, when the eigenvalue outcome is obtained at
$t=-1$\footnote{We take the time difference in integer values for
convenience}, and the eigenvalue of $\pm 1$ would be obtained.
Therefore, we see that if we take the result in \cite{song1}, the
eigenvalue outcome (at $t=-1$) takes place prior to the basis
choice at $t=0$.  This justifies our method of considering the
physical reality of $N_Q$ as a function of $N_C$ if we assume the
claim in \cite{song1} is correct.

Let us suppose that two spatially distant parties, called Alice
and Bob, share maximally entangled qubits $Q_1$ and $Q_2$, as
follows:
\begin{equation}
|\Psi\rangle = \frac{1}{\sqrt 2}(|0\rangle_{Q_1} |0\rangle_{Q_2} -
|1\rangle_{Q_1} |1\rangle_{Q_2}). \label{max}\end{equation} Now
suppose Alice and Bob choose to measure either in $Z$ or $X$ basis
with $P=1/2$, respectively, and the eigenvalue outcomes $\pm 1$
are obtained at each end.  Let us focus our attention to two
events taking place at two different times, i.e., when the basis
choice is made which we will set to be at $t=0$ and when the
eigenvalue outcome is obtained at $t=-1$. That is, the eigenvalue
outcome is a prior event than the basis choice and we wish to
examine correlation of basis choices with respect to the
eigenvalue outcomes.  Deutsch and Hayden gave a useful notation in
describing the entangled qubits in the Heisenberg picture.
However, for our purpose, it would be more convenient to discuss
the correlation of basis choices with respect to the eigenvalue
outcomes in the Schr\"odinger picture. Although in \cite{song1},
it was shown that the Heisenberg picture that gives a correct
picture rather than the Schr\"odinger picture, this only applies
to a very special case, just as Newton's theory is much more
convenient and sufficient method in most cases even though
relativity is the correct description.  Since it would more
convenient in analyzing the basis choices with respect to
eigenvalue, we wish to write down the process in the Schr\"odinger
picture. The only thing to remember is that we will consider the
basis choice with respect to the eigenvalue outcomes.

We will assume Alice and Bob each possess extra qubits $Q_3$ and
$Q_4$, which will be used to represent their choice of measurement
basis, respectively. Therefore, Alice has $Q_1$ and $Q_3$, while Bob
contains $Q_2$ and $Q_4$ at his end. Now, Alice and Bob can measure
$Q_1$ and $Q_2$ in either $Z$ or $X$ bases with $P=1/2$,
respectively. If Alice chooses to measure $Q_1$ in $Z$ basis, she
prepares $Q_3$ to be $|0\rangle_{Q_3}$ and does not apply any
unitary operation on $Q_1$. If Alice wants to measure in $X$ basis,
she then prepares $|1\rangle_{Q_3}$ and applies the Hadamard
operation, $H$, to $Q_1$. Therefore, the outcome of $Q_3$ will
indicate the chosen basis while the outcome of $Q_1$ will be either
$+1$ or $-1$.  It should be noted that we will consider $Q_3$ and
$Q_4$ represent observers' choice of observables.  Following the
suggestion in \cite{song1}, we will consider $Q_3$ and $Q_4$ as
functions of $Q_1$ and $Q_2$.

Alice could choose $Z$ or $X$ by preparing $Q_3$ to be in
$\frac{1}{\sqrt 2}(|0\rangle +|1\rangle )_{Q_3}$ initially, then
projecting it onto $\{ |0\rangle ,|1\rangle\}$.  If the outcome is
$+1$, then she doesn't apply any unitary operation on $Q_1$, while
Hadamard gate is applied for $-1$ outcome of $Q_3$. Or Alice can
toss a coin\footnote{We assume the coin used by Alice had no
previous contact with the coin used at Bob's end.} and prepare
$|0\rangle_{Q_3}$ for heads while preparing $|1\rangle_{Q_3}$ and
applying Hadamard to $Q_1$ for tails. Likewise, Bob can prepare
$|0\rangle_{Q_4}$ if he wishes to measure in $Z$ while preparing
$|1\rangle_{Q_4}$, and apply $H$ to $Q_2$ when it is to be measured
in $X$ basis. The final state for qubits $Q_1,Q_2,Q_3$, and $Q_4$
can then be written as follows:
\begin{eqnarray}
\rho &=& \frac{1}{4} |0\rangle_{Q_1}\langle 0| \otimes
|0\rangle_{Q_2}\langle 0| \otimes  \Big[ \frac{1}{2}
|0\rangle_{Q_3}\langle 0| \otimes |0\rangle_{Q_4}\langle 0| \nonumber \\
& &\; +\; \frac{1}{4} |0\rangle_{Q_3}\langle 0| \otimes
|1\rangle_{Q_4}\langle 1|
+\frac{1}{4} |1\rangle_{Q_3}\langle 1| \otimes |0\rangle_{Q_4}\langle 0| \Big]  \nonumber \\
 & + & \frac{1}{4} |0\rangle_{Q_1}\langle 0| \otimes  |1\rangle_{Q_2}\langle 1| \otimes
\Big[ \frac{1}{2} |1\rangle_{Q_3}\langle 1| \otimes |1\rangle_{Q_4}\langle 1| \nonumber \\
& &\; + \; \frac{1}{4} |0\rangle_{Q_3}\langle 0| \otimes
|1\rangle_{Q_4}\langle 1|
+\frac{1}{4} |1\rangle_{Q_3}\langle 1| \otimes |0\rangle_{Q_4}\langle 0|\Big] \nonumber \\
& + & \frac{1}{4} |1\rangle_{Q_1}\langle 1| \otimes
|0\rangle_{Q_2}\langle 0| \otimes
 \Big[ \frac{1}{2} |1\rangle_{Q_3}\langle 1| \otimes |1\rangle_{Q_4}\langle 1| \nonumber \\
&  & \; + \; \frac{1}{4} |0\rangle_{Q_3}\langle 0| \otimes
|1\rangle_{Q_4}\langle 1|
+\frac{1}{4} |1\rangle_{Q_3}\langle 1| \otimes |0\rangle_{Q_4}\langle 0|\Big]  \nonumber \\
& + & \frac{1}{4} |1\rangle_{Q_1}\langle 1| \otimes
|1\rangle_{Q_2}\langle 1| \otimes
\Big[ \frac{1}{2} |0\rangle_{Q_3}\langle 0| \otimes |0\rangle_{Q_4}\langle 0| \nonumber \\
&  & \; + \; \frac{1}{4} |0\rangle_{Q_3}\langle 0| \otimes
|1\rangle_{Q_4}\langle 1|
+\frac{1}{4} |1\rangle_{Q_3}\langle 1| \otimes |0\rangle_{Q_4}\langle 0|\Big]  \nonumber \\
&+& {\rm {off-diagonal\; terms}}. \label{rho}\end{eqnarray} We now
discuss the state $\rho$ in (\ref{rho}) in terms of elements of
reality. By following the logic structure used in Hardy's proof
\cite{hardy}, with the state $\rho$ in (\ref{rho}), we would like to
consider elements of physical reality for $Q_3$ and $Q_4$, which
represent the basis choice, as a function of $Q_1$ and $Q_2$. Now,
let us make an assumption that an element of reality of $Q_3$
($Q_4$) will not be a function of $Q_2$ ($Q_1$). The probability of
obtaining $+1$ and $+1$ for $Q_3$ and $Q_4$ when $Q_1=+1$ and
$Q_2=+1$ is 1/2. In the following, let us consider these 50\% of the
runs where $Q_3$ and $Q_4$ will both be $+1$ when $Q_1=+1$ and
$Q_2=+1$. According to (\ref{rho}), when $+1$ and $+1$ are obtained
from $Q_1$ and $Q_3$, we can predict with certainty the outcome of
$Q_4$ will be $-1$ as long as the outcome of $Q_2$ is $-1$.
Therefore, according to the reality criterion given by EPR the value
for the elements of reality for the outcome of $Q_4$ as a function
of $Q_2$ is the following,
\begin{equation}
N_{Q_4}(N_{Q_2}=-1) = -1. \label{value1}\end{equation} In the 50\%
of the runs we are considering, $N_{Q_4}$ in (\ref{value1}) should
not change even if $Q_1$ is $-1$ instead of $+1$ since we assumed
the element of reality for $Q_4$ will not be a function of $Q_1$.
Similarly, in these runs, when the outcomes of $Q_2$ and $Q_4$ are
$+1$ and $+1$, respectively, when $Q_1=-1$, $Q_3$ must yield $-1$.
Therefore, we can deduce the value of the element of reality for
$Q_3$ with respect to $Q_1$ in these events as
\begin{equation}
N_{Q_3}(N_{Q_1}=-1) = -1. \label{value2}
\end{equation}
Since we derived the value for the element of reality for $Q_3$
with no dependence on $Q_2$, these values should not change when
$Q_2=-1$. Along with $N_{Q_4}$'s independence of $Q_1$, this
contradicts the fact that, according to (\ref{rho}), when $Q_1=-1$
and $Q_2=-1$, the probability of getting $-1$ and $-1$ for $Q_3$
and $Q_4$ is zero. Therefore, we are led to an inconsistency when
we assumed an element of reality for $Q_3$ ($Q_4$) cannot be a
function of $Q_2$ ($Q_1$)'s outcome. If the value for the element
of reality for $Q_3$ can be derived not only as a function of
$Q_1$ but also of $Q_2$, and similarly for $Q_4$ as a function of
both $Q_1$ and $Q_2$, the contradiction can be avoided. While
using the logic in \cite{hardy}, we've shown that an element of
reality for $Q_3$ is a function of at least $Q_1$ and $Q_2$ while
$Q_4$'s reality is also dependent on the same $Q_1$ and $Q_2$.
Therefore, we conclude that, with respect to the outcome of
entangled quantum systems $Q_1$ and $Q_2$, $Q_3$ and $Q_4$ cannot
be independent from each other.

We have shown that if we take the claim in \cite{song1}, i.e., the
observer's reference frame going backwards in time, the assumption
in Bell-type proofs, which states that the basis choice at two
ends are not correlated at two ends, may not be valid. Certainly,
there is always the possibility that two detectors might have
interacted a long time ago.  However, what we have shown in
regards to the correlation of measurement bases is fundamentally
different from such possibility. We considered that the basis
between $X$ and $Z$ may be chosen by collapsing a superposed
qubit.  Since this qubit is a pure state, there should not be any
correlation with any other system, including a detector, at the
other end.  Even when the basis was chosen randomly based on
quantum probability, we have shown that there is still a
possibility that two basis choices are correlated.


\section*{Acknowledgments}
I thank Manny Knill for his valuable suggestions for many
improvements on an earlier version of this paper and Gavin Brennen
for helpful discussions.  I am also grateful to the referees for
valuable suggestions.



\begin{thebibliography}{00}
\bibitem{bell} J.S. Bell, Physics {\bf {1}}, 195 (1964).

\bibitem{CHSH} J.F. Clauser, M.A. Horne, A. Shimony, and R.A. Holt,
Phys. Rev. Lett. {\bf{23}}, 880 (1969).

\bibitem{GHZ} D.M. Greenberger, M.A. Horne, and A. Zeilinger, in {\it {Bell's Theorem, Quantum Theory, and Conception of the
Universe}} ed. M. Kafatos (Dordrecht, Kluwer 1989); D.M.
Greenberger, M. Horne, A. Shimony, and A. Zeilinger, Am. J. Phys.
{\bf {58}}, 1131 (1990).

\bibitem{hardy} L. Hardy, Phys. Rev. Lett. {\bf {71}}, 1665 (1993).

\bibitem{aspect} A. Aspect, J. Dalibard, and G. Roger, Phys. Rev. Lett. {\bf {49}}, 1804 (1982).

\bibitem{adami} N.J. Cerf and C. Adami, Phys. Rev. Lett. {\bf{79}}, 5194 (1997).


\bibitem{winter} M. Horodecki, J. Oppenheim, and A. Winter, Nature
{\bf{436}}, 673 (2005).

\bibitem{hayden2} P. Hayden, Nature {\bf{436}}, 633 (2005)





\bibitem{EPR} A. Einstein, B. Podolsky, and N. Rosen, Phys. Rev. {\bf 47}, 777 (1935).



\bibitem{song1} D. Song, Int. J. Theor. Phys. (in press) (2007); arXiv:quant-ph/0610047v1.

\bibitem{song2} D. Song, NeuroQuantology {\bf{5}}, 382 (2007).











\end{thebibliography}
\end{document}